\documentclass[twocolumn]{aastex631}
\usepackage{amsmath}
\usepackage{graphicx}
\usepackage{braket}
\usepackage{enumitem}
\usepackage[dvipsnames]{xcolor}

\renewcommand{\>}{\rangle}
\newcommand{\Msun}{M_{\odot}}
\newcommand{\Lenstool}{\text{\sc{Lenstool}}}
\newcommand{\Ls}{L_*}
\newcommand{\rts}{r_{\text{t}*}}
\newcommand{\eee}{\equiv}
\newcommand{\rmd}{{\rm d}}

\newcommand {\sref}[1]        {Section~\ref{#1}}
\newcommand {\aref}[1]        {Appendix~\ref{#1}}
\newcommand {\fref}[1]        {Figure~\ref{#1}}
\newcommand {\tref}[1]        {Table~\ref{#1}}
\newcommand {\eref}[1]        {Equation~(\ref{#1})}

\shorttitle{New CDM Crisis}
\shortauthors{Natarajan, Chiang, \& Dutra}

\begin{document}

\title{New Cold Dark Matter Crisis Revealed by Multiscale Cluster Lensing}

\author[0000-0002-5554-8896]{Priyamvada Natarajan}
\affiliation{Department of Astronomy, Yale University, New Haven, CT 06511, USA}
\affiliation{Department of Physics, Yale University, New Haven, CT 06511, USA}
\email{priyamvada.natarajan@yale.edu}

\author[0000-0002-9370-4490]{Barry T. Chiang}
\affiliation{Department of Astronomy, Yale University, New Haven, CT 06511, USA}

\author[0000-0001-7040-4930]{Isaque Dutra}
\affiliation{Department of Physics, Yale University, New Haven, CT 06511, USA}


\begin{abstract}
The properties of substructure in galaxy clusters, exquisitely probed by gravitational lensing, offer a stringent test of dark matter (DM) models. Combining strong- and weak-lensing data for massive clusters, we map their total mass—dominated by DM—over the dynamic range needed to confront small-scale predictions for collisionless cold DM (CDM). Using state-of-the-art lens models, we extract four key subhalo properties: the mass function, projected radial distribution, internal density profile, and tidal truncation radius. We find that the subhalo mass functions and truncation radii are consistent with CDM expectations. In contrast, the inner density profiles and radial distributions of subhalos are strongly discrepant with CDM. The incidence of galaxy--galaxy strong lensing from subhalo cores exceeds CDM predictions by nearly an order of magnitude, requiring inner density slopes as steep as $\gamma \gtrsim 2.5$ within $r \lesssim 0.01R_{200}$ consistent with core-collapsed self-interacting DM (SIDM), while the same subhalos behave as collisionless in their outskirts. Additionally, the observed radial distribution of subhalos hosting bright cluster member galaxies, explicitly modeled in the lens reconstructions, remains incompatible with CDM. Taken together, these small-scale stress tests reveal an intriguing paradox and challenge the DM microphysics of purely collisionless CDM, motivating hybrid scenarios---such as a dual-component model with both CDM and SIDM or entirely new classes of DM theories.
\end{abstract}


\section{Introduction}

The Lambda cold dark matter  ($\Lambda$CDM) paradigm excels in accounting for observations on large scales, matching data from the cosmic microwave background, baryon acoustic oscillations, and the derived matter power spectrum. However, persistent small-scale tensions remain \citep{DelPopolo2017Galax517D}. Many of these tensions---including abundance mismatches, for instance, with the missing satellites problem, and internal structure issues with the density profile shape leading to the cusp--core problem and the dwarf galaxy diversity problem---have also been largely resolved with more sophisticated and improved simulation suites \citep{Bullock2017,Sales+2022,Chiang2025,Cruz+2025}.

On massive lensing cluster scales, however, given their rarity and high masses, it has been challenging to find mass-matched simulated samples for cluster lenses \citep{PNSpringel2004,PNDLSpringel2007,Natarajan2017}. The availability of TNG-Cluster (hereafter, TNG-C; \citealt{Nelson2024}), which resimulates $\sim$350 clusters drawn from an extremely large 1 ${\rm Gpc}^3$ volume (36 times larger than the previous state-of-the-art TNG300 simulation box), has been transformative for comparing detailed substructure properties. TNG-C is the first simulation suite that provides apt mass-matched samples for the comparison of properties with observationally inferred subhalos inside massive lensing clusters over cosmic time.

In this work, we confront substructure properties derived from observed cluster lenses with those derived from mass-matched CDM cluster analogs from the TNG-C simulations. In doing so, we demonstrate here that a particularly sharp, hitherto undocumented, and now data-rich challenge for CDM on small scales arises in massive-cluster lenses. There are four small-scale substructure diagnostics that we explore here: the subhalo mass function (SHMF), radial distribution, internal structure, and outer truncation radii for comparison with simulations. The emerging picture, taking all the results together, poses an intriguing set of crises for CDM that warrants stress testing the model.

The outline of this Letter is as follows. We first describe the observed cluster sample (\sref{sec:obs_cluster}) and lens model construction methodology (\sref{sec:lens_models}), as well as the properties of the corresponding simulated cluster analogs (\sref{sec:sim_tngc}). Next, we compare each of the following diagnostics of the observed substructure properties against theoretical predictions from simulated CDM analogs in TNG-C: the SHMF (\sref{sec:shmf}), projected radial distributions (\sref{sec:radial_dist}), inner density profiles (\sref{sec:ggsl_tension}), and outer truncation radii (\sref{sec:trun_radii}). We discuss the sources of uncertainty in our analysis in \sref{sec:rt_Uncertainties}. We conclude in \sref{sec:conclusions} with a discussion of cluster-scale substructure properties and their implications for DM microphysics.

\section{Observed Cluster Sample}\label{sec:obs_cluster}

In this work, we focus on the study of three massive galaxy clusters that span $z=0.39$\textendash$0.54$, drawn from the observed Hubble Space Telescope (HST) Frontier Fields \citep{HSTFrontierFieldsData} and CLASH \citep{Balestra2016ApJS22433B} programs. As listed in \tref{tab:Cluster_Properties}, this sample was chosen for the publicly available extensive ground-based spectroscopic follow-up data, and hence it has some of the best-constrained lensing mass models at present.

\begin{table*}[ht]
        \caption{Galaxy Cluster Sample\label{tab:Cluster_Properties}}
        \begin{tabular}{cccccccc}
			\hline Cluster
			& $\<z_\text{spec}\>$ & R.A. (deg) & Decl. (deg) & $M_{200}$ ($\Msun$) & $R_{200}$ (Mpc)  & $N^\text{\rm spec}_{\rm gal}$ &References\\ 	\hline 	\hline 
			MACS J0416 &  0.3972  & 64.0381   & $-24.0675$   &  $1.53\times10^{15}$  &   2.69 &  66  &  (1), (2), (3) \\	 
			MACS J1206 &  0.4398  & 181.5506 &   $-8.8009$  &   $1.37\times10^{15}$  &  1.96  &  152 &  (4), (5), (6), (7) \\ 
			MACS J1149 &  0.5420  &  177.3990  &  22.3979  &  $1.27\times10^{15}$  &  1.84  &  144     &  (3), (8) \\	 \hline
		\end{tabular}
        \tablecomments{From left to right, we show the respective mean spectroscopic redshift $\<z_\text{spec}\>$, R.A. (J2000), decl. (J2000), best-fit NFW virial halo mass $M_{200}$, virial radius $R_{200}$, and number of spectroscopically confirmed and \Lenstool-identified member galaxies $N^\text{\rm spec}_{\rm gal}$.}
\tablerefs{(1) \citet{Balestra2016ApJS22433B}, (2) \citet{Umetsu2016ApJ821116U}, (3) \citet{Lotz2017ApJ83797L}, (4) \citet{Biviano2023ApJ958148B}, (5) \citet{Biviano:2013eia}, (6) \citet{Umetsu2012ApJ75556U}, (7) \citet{Bergamini2019AA631A130B}, (8) \citet{Grillo2016ApJ82278G}.}
	\label{tab:Cluster_Properties}
\end{table*}

\begin{enumerate}
\item MACS J0416.1-2403: appears dynamically relaxed \citep{Postman2012ApJS19925P}. We adopt the \Lenstool-optimized cluster lensing model from the HST Frontier
Fields Initiative data \citep{Lotz2017ApJ83797L, Natarajan2017} and independently identify cluster members
using the CLASH galaxy catalog (see \citealt{Chiang2026a} for details).
\item MACS J1206.2-0847: appears to be a relaxed cool-core
cluster \citep{Ebeling2009MNRAS3951213E, Postman2012ApJS19925P, Girardi:2015aga}. We adopt the \Lenstool-optimized lensing model of \citet{Caminha2017AA607A93C} and again identify cluster members using the CLASH
galaxy catalog (see \citealt{Chiang2026a}).
\item MACS J1149.6+2223: appears dynamically relaxed \citep{Postman2012ApJS19925P, Finney2018ApJ85958F}. We adopt the \Lenstool-optimized lensing model from the HST Frontier Fields Initiative data and crossmatch cluster member galaxies with spectroscopic information from the SIMBAD database \citep{Wenger2000AAS1439W}.
\end{enumerate}

\section{Summary of cluster lens modeling}\label{sec:lens_models}

We briefly describe the parametric mass models derived by combining strong- and weak-lensing observations of cluster lenses. This modeling methodology adopts self-similar parametric profiles for cluster members, providing constraints on the structural properties of subhalos hosting cluster member galaxies. Parametric mass models are chosen as they are best suited for the direct comparison of lensing-inferred subhalo properties with CDM simulations and their halo catalogs; their utility and suitability for this purpose is well established in the literature (see the review by \cite{Natarajan2024}).

The publicly available software package \Lenstool\footnote{\url{https://projets.lam.fr/projects/lenstool/wiki}} permits the use of observed lensing signals, including the positions and brightnesses of the multiply imaged strongly lensed galaxies, as well as the positions and shapes of the weakly lensed background galaxies, to reconstruct the detailed mass distribution in massive lensing clusters. The methods used here are standard and have been in use for over two decades, having been amply tested against multiple independent cosmological simulation suites. Reviews of these modeling methods, along with their powers and limitations, can be found in \cite{Natarajan2024,Kneib2011A&ARv1947K}. \Lenstool~performs multiscale Bayesian optimization on joint strong+weak-lensing constraints to construct mass distributions modeled as the linear superposition of parametric profiles on a range of scales, as proposed by the conceptual
model presented in \citet{Natarajan1997MNRAS287833N,Jullo2007,Niemiec2020}:
\begin{align}
	\phi_\text{tot} = \sum_i \phi^\text{halo}_i + \sum_j \phi^\text{subhalo}_j +  \phi_\kappa, 
\end{align}
where $\phi^\text{halo}_i$ represents the Mpc-scale halos associated with the smooth large-scale cluster gravitational potentials; $\phi^\text{subhalo}_j$ are the kiloparsec-scale subhalos associated with cluster member galaxies; and $\phi_\kappa$ is a potential constant external shear field. Further details of the mass modeling are provided in \aref{app:lens_modeling}.

For each observed cluster, the subhalo sample is constructed by selecting the brightest cluster member galaxies based on the HST F814W ($I$-band) luminosity, associating each with a subhalo derived from established scaling relations (\eref{eqn:E_Scaling_Relations}). This procedure associates subhalos with the entire population of cluster member galaxies down to a luminosity cut and is agnostic to the specific observed small-scale galaxy–galaxy strong-lensing (GGSL) features; instead, the entire multiscale lens models are simultaneously optimized, assuming light traces mass using the available strong- and weak-lensing constraints. Detailed membership identification of the spectroscopically confirmed cluster galaxies can be found in Appendix~A of \citet{Chiang2026a}. Here, we treat the lensing-inferred total mass associated with each subhalo $M_\text{dPIE}$, defined in \eref{eqn:rho_dPIE}, as a proxy for the subhalo total mass $M_\text{sub}$, which for simulated subhalos means the mass of all gravitationally bound particles. As shown in \citet{Meneghetti2017MNRAS4723177M}, the \Lenstool-optimized $M_\text{dPIE}$ statistically and unbiasedly recovers $M_\text{sub}$ as measured in cluster-scale simulations. This method allows for a robust comparison between observationally inferred and simulated subhalo properties.

\section{The TNG-Cluster simulation: apt comparator for cluster lenses}\label{sec:sim_tngc}

To find the closest redshift- and mass-matched simulated analogs to our cluster sample, we use the TNG-C simulations\footnote{\url{https://www.tng-project.org/cluster/}}---a next-generation extension of the IllustrisTNG project, designed specifically to model a large sample of galaxy clusters at high resolution in a fully cosmological context \citep{Nelson2024}. The suite comprises zoom-in magnetohydrodynamical simulations of 352 massive galaxy clusters over the mass range $M_{\rm 200} = 10^{14\textendash15.5}\,\Msun$, run with the moving-mesh code \textsc{Arepo} \citep{Springel2010MNRAS401791S} and the full IllustrisTNG galaxy formation model \citep{Pillepich2018MNRASb}. To preserve the large-scale cosmological environments, target clusters are selected from large DM-only parent boxes and then resimulated at higher resolution, reaching a fixed DM (baryon) particle mass resolution of $6.1\times10^7$ ($1.2\times10^7$)~$\Msun$ and a gravitational softening length of $1.48$~kpc. The TNG-C public catalog provides subhalos identified by \textsc{SubFind} \citep{Springel2001}, the limitations of which are well known. We discuss these details in \aref{app:num_rad_sub} and show that this choice does not impact the findings presented in this work.

In addition to the primary aim of studying the baryon cycle in clusters (Intra-Cluster Medium thermodynamics, metal enrichment, and cool cores/noncool cores), the TNG-C suite is also ideal for mass-matched comparisons with massive-cluster lenses and their substructure demographics. This suite provides the first robust CDM predictions, with fully coupled baryonic physics, for SHMFs, their inner density profiles, and tidal truncation radii in massive galaxy clusters, against which the lensing-inferred properties of observed massive clusters can be calibrated and directly compared.

To select the most representative simulated analogs, we search in the closest redshift-matched data outputs of our sample\textemdash MACS~J0416 (Snapshot \#72), MACS~J1206 (Snapshot \#70), and MACS~J1149 (Snapshot \#65)\textemdash and find the five best-mass-matched systems per cluster, giving $\Delta M_{200}/M_{200} = 0.010$\textendash$0.089$ at the end. To self-consistently model the observational selection functions in assigning subhalos, we use the member galaxy properties associated with all bound subhalos of a simulated cluster, apply rest-frame $V$-band luminosity cuts\footnote{The observational selection is based on the HST F814W ($I$-band), which at the cluster redshifts of our sample ($z \approx 0.39$\textendash$0.54$) corresponds to a rest-frame wavelength of $\lambda_{\rm rest} \approx 5200$\textendash$5800$\,\AA. This range is best traced by the rest-frame $V$ band ($\lambda_{\rm eff} \approx 5500$\,\AA) provided in the TNG-C core catalogs \citep{Nelson2024}, which are computed using \citet{Bruzual2003} stellar population synthesis models.} equivalent to the observed $I$-band apparent magnitude limits at the corresponding redshifts adopted for the data ($m = 27.1$\textendash$18.7$ for MACS~J0416; $m=26.4$\textendash$14.9$ for MACS~J1206; $m =22.0$\textendash$16.7$ for MACS~J1149), and select the subhalos hosting the $N^\text{\rm spec}_{\rm gal}$-brightest member galaxies as the spectroscopically confirmed analogs\footnote{We have explicitly verified that the results presented in this work remain unchanged under alternative subhalo-mass-based selection criteria for assigning subhalo analogs.}. Such selected subhalos in simulated cluster analogs are found to produce galaxy luminosity functions consistent with the respective observed distributions. Hence, the simulated analogs comprise the same number of subhalos as deployed in each observed cluster lens mass model (\tref{tab:Cluster_Properties}) and provide the best-fit combined mass distribution of both large- and small-scale components.

\section{Subhalo mass function}\label{sec:shmf}

\begin{figure}
\begin{minipage}{0.48\textwidth} 
        \includegraphics[width=\textwidth]{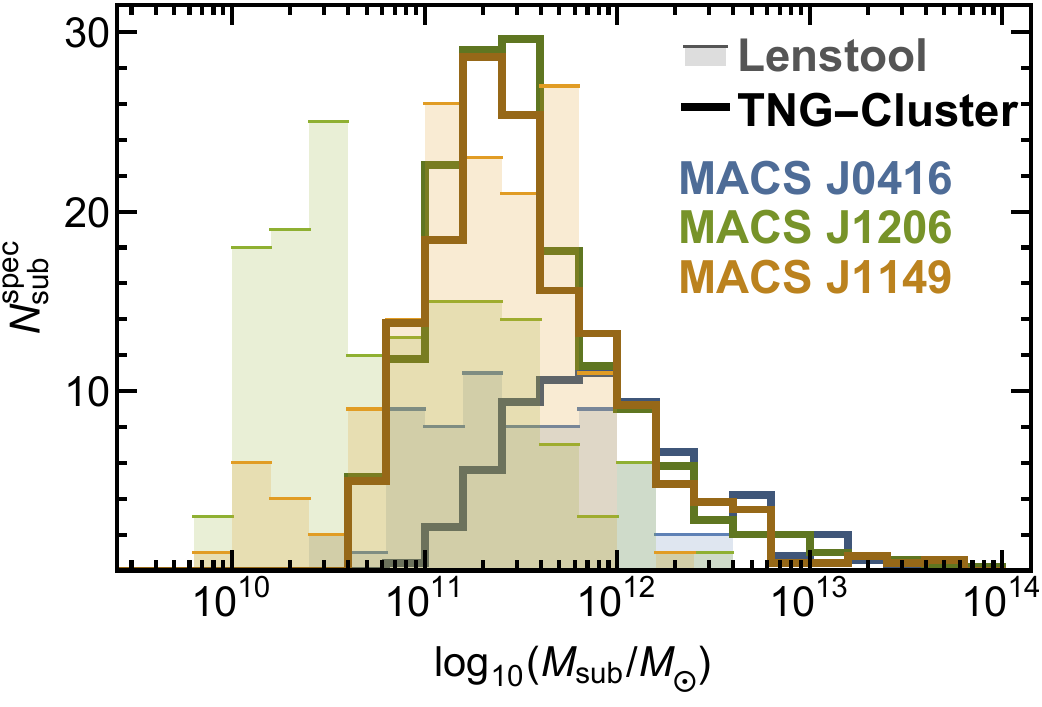}
\caption{The mass function of subhalos associated with spectroscopically confirmed bright cluster member galaxies in observed lensing clusters (\Lenstool, color-shaded) compared with CDM predictions from simulated analogs (TNG-C, solid lines) under the same selection criteria; see \sref{sec:sim_tngc} for details.}
\label{fig:SHMF.pdf}
\end{minipage}
\end{figure}

Structures assemble hierarchically in CDM, as smaller halos merge to form more massive structures at later times. The mass function of substructures is a robust prediction of the model, with a power-law slope of $dn/dM_\text{sub} \propto M_\text{sub}^{-1.9}$ \citep[e.g.,][]{vandenBosch2016MNRAS}. The SHMF inside collapsed structures with masses corresponding to those of observed cluster lenses $M_{200} \sim 10^{15} \Msun$ can be computed directly from the simulated analogs. In this work, as noted above, we partition the total mass of a cluster into large-scale mass components and smaller-scale subhalos that are associated with bright cluster member galaxies. 

\fref{fig:SHMF.pdf} compares the SHMFs associated with the spectroscopically confirmed member galaxies of each observed cluster with those of their simulated analogs (averaged over five best-matched systems per cluster, to take into account and average over orientations and ellipticities). We note reasonably good agreement between the lensing-derived values and CDM predictions, except for MACS J1206, where simulations cannot account for the low-mass end of the observed bimodal distribution. We attribute this mismatch to the fact that even in the TNG-C volume, we do not find an analog of MACS J1206 that appears as strongly elongated within the central $\sim200$~kpc in projection \citep[e.g.,][]{Caminha2017AA607A93C}.

\section{Radial Distribution of Substructure}\label{sec:radial_dist}

\begin{figure}
    \includegraphics[width=0.99\linewidth]{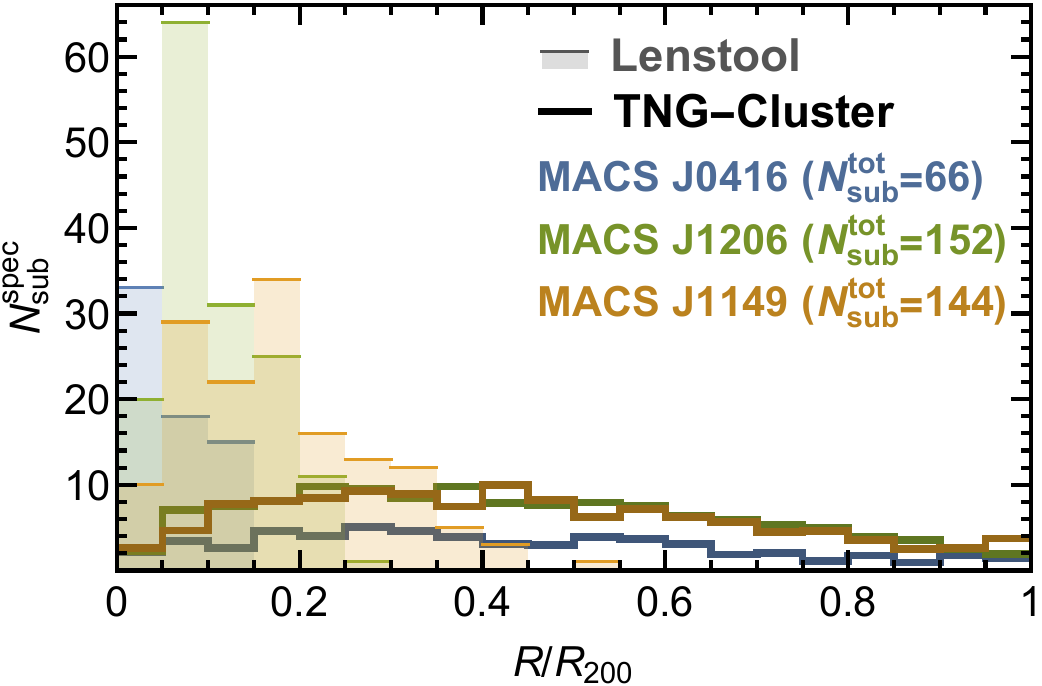}
    \hfill
    \includegraphics[width=0.99\linewidth]{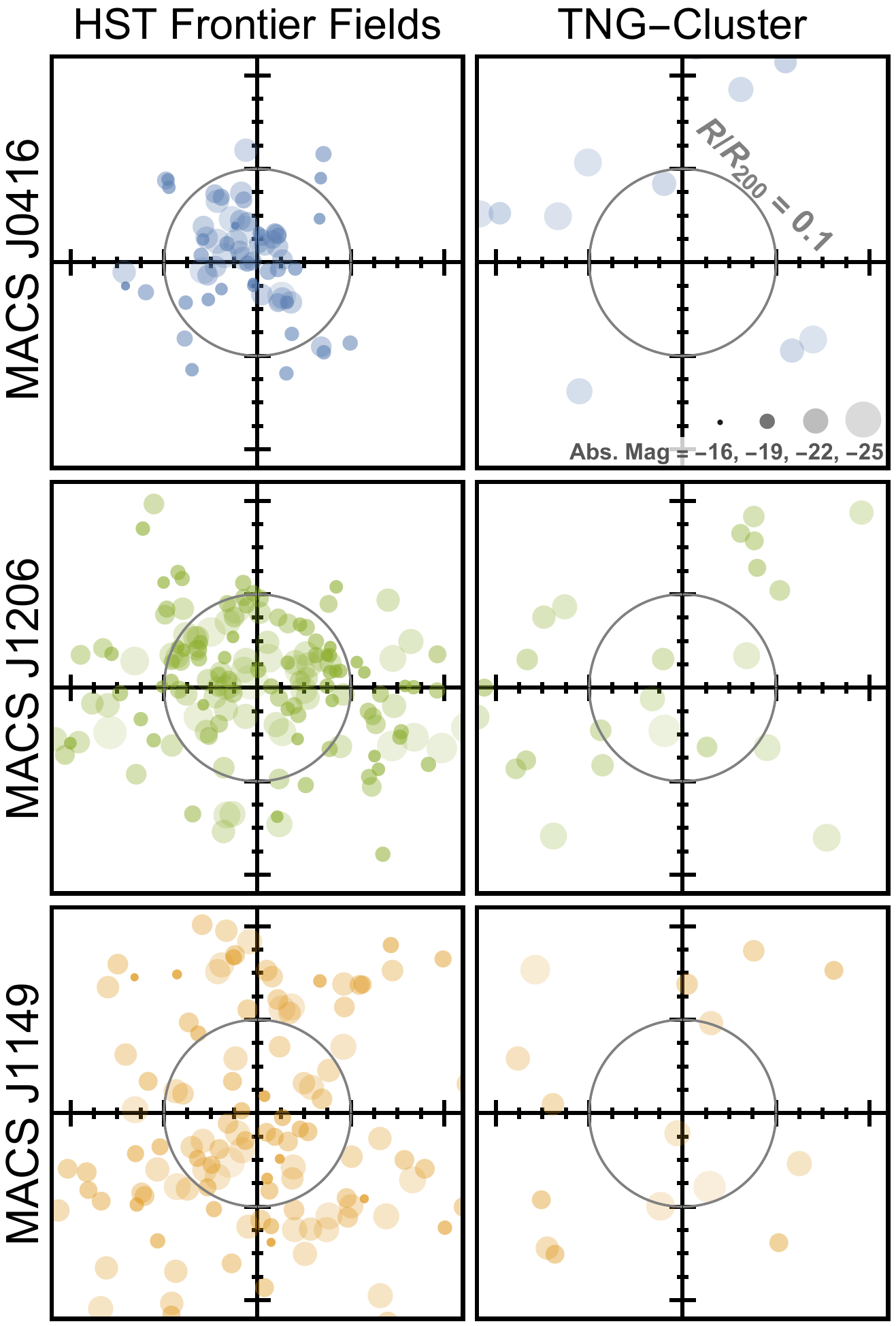}
	\caption{Top: the projected radial distribution of subhalos associated with spectroscopically confirmed member galaxies in observed clusters (color-shaded) and in simulated analogs (solid lines; averaged over the five best-matched analogs and over three random projections). Bottom: the projected spatial distribution of subhalos around cluster centers in observations (left) and simulated analogs (right; from the best-mass-matched analog along a random projection), annotated by the respective absolute magnitude of associated galaxies. As compared to the observations, the simulated analogs clearly show a dearth of and fail to reproduce the spatial clustering of inner substructures within $R/R_{200}\lesssim 0.2$.}
	\label{fig:combined_figures}
\end{figure}

For each observed cluster lens in our sample, we next compare the projected radial distribution of subhalos that host observed cluster member galaxies and their similarly selected simulated TNG-C analogs, as shown in the top panel of \fref{fig:combined_figures}. Across all three systems, there is a persistent and stark inconsistency between observations and simulations, where the latter exhibit a dearth of substructures at small projected halocentric radii $R/R_{200}\lesssim 0.2$ and instead appear to have a roughly radius-independent distribution. To further illustrate this discrepancy, the bottom panels of \fref{fig:combined_figures} compare the projected 2D distributions of observations and simulations. We observe that the simulated cluster analogs fail to reproduce not only the abundance but also the spatial clustering of these substructures at small projected halocentric radii.

The stark dearth of inner substructures in simulated CDM clusters was first hinted at in the analysis of the cluster Cl~0024+16 by \citet{Natarajan2009ApJ693970N}, and formally codified in \citet{Natarajan2017} with the subsequent detailed analysis of three additional HST Frontier Fields clusters. The authors noted that this discrepancy could also be partly caused by numerical artifacts\textemdash the inaccurate numerical modeling of dynamical friction and tidal stripping or limited accuracy in the subhalo finding algorithm\textemdash leading to an artificial reduction of inner substructures in simulated clusters. Notably, this mismatch has also been observed on the Milky Way mass scales \citep[e.g.,][]{Campbel2018, Graus2019, Carlsten2020}, where the radial distribution of satellites disagrees with CDM analogs, in the same manner as we report here on cluster scales. This is noteworthy, as the same disagreement is independently reported on two different mass scales and in distinct environments, pointing to the same underlying\textemdash perhaps endemic\textemdash issue within CDM.

To probe the impact of potential numerical artifacts, \citet{Chiang2026Universal} have recently demonstrated that in
state-of-the-art cosmological simulations with kiloparsec-scale force resolution, subhalos with orbital pericenter distances $\lesssim0.2R_{200}$ are nearly all force-unresolved, leading to artificially enhanced tidal mass loss and runaway disruption \citep{vandenBosch2018MNRAS4754066V}. In parallel, recent halo finder comparison studies by \citet{Mansfield2024} and \citet{ForouharMoreno2025} have demonstrated
that conventional subhalo finders that rely only on the spatial clustering of particles\textemdash such as \textsc{SubFind} used in the TNG-C suite here\textemdash are particularly prone to losing track of subhalos at small pericentric distances, an issue that can be addressed by additionally accessing velocity-space information or tracking particles across simulation outputs. Indeed, when inspecting the 3D halocentric distances of these simulated subhalos, we find that the number counts are roughly constant between $0.3$ and $1 R_{200}$ but drop precipitously with decreasing radii below $\lesssim 0.3 R_{200}$. Furthermore, when averaging over all simulated cluster analogs, only 10 simulated subhalos lie within a 3D radius of $\leq 0.2R_{200}$ under the luminosity-based selection adopted here, which is an order of magnitude smaller than $N^\text{spec}_\text{tot} \sim 100$ in observed clusters. 

To address these numerical uncertainties due to inadequate resolution, subhalo finder choices, and specific feedback models \citep[e.g.,][]{Despali2017}, we demonstrate in \aref{app:num_rad_sub} that the same discrepancy persists even after calibrating the inner subhalo abundance to a DM-only benchmark, by ``explicitly injecting'' these disrupted subhalos into the TNG-C cluster analogs. With 1 million bootstrapping iterations, we find that within projected inner radii bins $\leq 0.2R_{200}$, the
CDM predictions remain statistically discrepant with the observed abundances by $5\sigma$\textendash$40\sigma$ (\fref{fig:App_B}). Aside from the discrepant abundances, the observed inner substructures also show greater diversity in luminosity and more complex projected spatial clustering than the simulated counterparts. Therefore, even after taking into account the deficits produced by numerical artifacts, we find that the discrepancy persists.

\section{THE INNER DENSITY PROFILES OF SUBSTRUCTURES}\label{sec:ggsl_tension}

GGSL in massive clusters provides a sensitive probe of the internal structure of DM subhalos over $M_{\rm sub}\sim 10^{10}$--$10^{12}\,M_\odot$ \citep{Meneghetti2020, Meneghetti2023}. GGSL events quantify how efficiently cluster member galaxies, embedded in the smooth cluster potential, generate additional galaxy-scale strong-lensing regions beyond those produced by the cluster-scale critical curves alone \citep{Meneghetti2020, Meneghetti2023}. Operationally, the GGSL cross section $\sigma_{\rm GGSL}(z_s)$ is the total source-plane area at source redshift $z_s$ enclosed by the union of galaxy-scale caustics, and the corresponding probability is
\begin{equation}
P_{\rm GGSL}(z_s) \equiv \frac{\sigma_{\rm GGSL}(z_s)}{A_s},
\tag{2}
\end{equation}
where $A_s$ is the source-plane area mapped by the observational field of view \citep{Meneghetti2020}.

Using high-fidelity lens models of HST Frontier Fields and CLASH clusters, \citet{Meneghetti2020} showed that the $P_{\rm GGSL}$ derived from these observed cluster lenses exceed predictions from state-of-the-art hydrodynamical CDM simulations by more than an order of magnitude across a broad range of $z_s$. This tension persists after varying the numerical resolution, subgrid feedback, and line-of-sight projections \citep[e.g.,][]{Ragagnin2022, Meneghetti2023, Heinze2024}. Because lensing-inferred SHMFs remain broadly consistent with CDM expectations \citep{Natarajan2017, Dutra2025}, the discrepancy points primarily to the subhalo lensing efficiency and hence their \emph{inner} structure: member galaxies appear to occupy subhalos that are systematically more centrally concentrated than predicted by collisionless CDM simulations.

Two recent papers tighten this inference by: (1) exhausting the freedom available within collisionless CDM (including extreme baryonic rearrangements; \citealp{Tokayer2024}); and (2) mapping the required inner profile slopes to self-interacting DM (SIDM) core collapse \citep{Dutra2025}.

\subsection{Rearranging Mass within $\Lambda$CDM: Limits of Baryonic Solutions}

\citet{Tokayer2024} revisit the GGSL discrepancy using five clusters (AS1063, MACS~J0416, MACS~J1206, A2744, PSZ1~G311) with \Lenstool-based strong-lensing
mass models, confirming the elevated observed $P_{\rm GGSL}$. They then test whether redistributing mass \emph{within} the lensing-constrained subhalo truncation radius can reconcile the signal while preserving the total enclosed mass. Replacing dual pseudo-isothermal elliptical (dPIE) subhalos with truncated Navarro--Frenk--White (NFW)--like profiles does not resolve the discrepancy. Including explicit stellar components and applying adiabatic contraction \citep{Gnedin2004} boosts $P_{\rm GGSL}$ only by factors of a few, with the inner log slope saturating at $\gamma \lesssim 2$ on the relevant scales. They conclude that the GGSL discrepancy cannot be explained by missing baryonic physics within collisionless CDM \citep{Tokayer2024}.

\subsection{Gravothermal Core Collapse in SIDM as a Resolution?}

Building on this, \citet{Dutra2025} independently confirm the robustness of the discrepancy, by recalculating GGSL probabilities for the HST Frontier Fields clusters plus PSZ1~G311 using both parametric lens models and direct ray tracing through mass-matched IllustrisTNG analogs \citep{Nelson2019ComAC62N}, including realistic ellipticities and projection effects. The inclusion of these effects shifts $P_{\rm GGSL}$ to the $\lesssim 40\%$ level---still far below what is required.

To connect GGSL to microphysics, they adopt a generalized NFW profile for cluster subhalos:
\begin{equation}
\rho(r) = \rho_s \left(\frac{r}{r_s}\right)^{-\gamma}
\left[1+\left(\frac{r}{r_s}\right)^2\right]^{-(\beta-\gamma)/2},
\tag{3}
\end{equation}
where $\gamma$ is the inner slope, and $\beta$ is the outer slope. They study
the impact of varying $\gamma$ at fixed total mass and hold $\beta$ near its CDM value. Ray tracing through such modified subhalo populations yields a strong monotonic response: steepening from $\gamma=1$ to $\gamma\simeq 2.5$ increases $P_{\rm GGSL}$ by $\sim 3.5$ times, and to $\gamma\simeq 2.9$ by $\sim 8$ times, bringing the simulations substantially
closer to observations. They further show that the implied central density rises by $\sim 4$ orders of magnitude within $0.1\,r_{200}$, as expected for deeply core-collapsed systems.

Assessing whether known baryonic processes could generically yield $\gamma\gtrsim 2.5$ on kiloparsec scales in collisionless CDM---including DM spikes around supermassive black holes \citep{GondoloSilk1999} and ``mound'' configurations driven by black hole feedback \citep{Bertone2025}---they
conclude that while these can steepen profiles on parsec scales, they do not produce ubiquitous kiloparsec-scale $\gamma\gtrsim 2.5$ slopes, as required. By contrast, SIDM models undergoing gravothermal core collapse appear to naturally generate post-collapse
inner slopes $\gamma\sim 2.5$--3 in the densest environments \citep[e.g.,][]{Balberg2002,YangYu2021}. Taken together with \citet{Tokayer2024}, the GGSL
discrepancy cannot be resolved by plausible rearrangements of mass in collisionless CDM and instead points to self-interactions
in the dark sector \citep{Tokayer2024,Dutra2025}.

\section{OUTER-HALO CONSTRAINTS FROM CLUSTER SUBHALO TRUNCATION RADII}\label{sec:trun_radii}

If what is required are core-collapsed SIDM subhalos to account for the inner regions, then it is worth examining if the outskirts of these same subhalos are consistent with this altered nature of DM. \citet{Chiang2026a} exploit an orthogonal outer halo diagnostic by using the tidal truncation radii of galaxy-scale subhalos inferred from high-fidelity strong+weak-lensing mass models. The key result is that the subhalo spatial extents in eight clusters (A2218, 383, 963, 209, and 2390 and the sample introduced in \sref{sec:obs_cluster}) are statistically consistent with collisionless CDM but are in strong tension with the compact truncation radii expected in the strongly collisional SIDM regime required to drive gravothermal core collapse in cluster subhalos \citep[e.g.,][]{Dutra2025,Tokayer2024}.

Their analysis uses \Lenstool-optimized parametric models for eight clusters, decomposing each cluster into cluster-scale halos plus galaxy-scale subhalos hosting spectroscopically confirmed members. The subhalos are modeled as dPIE profiles, with the lensing-inferred dPIE truncation radius $r_{t,{\rm dPIE}}$ providing a directly measured proxy for the subhalo tidal extent.

For CDM, the predicted tidal radius $r^{\rm CDM}_{t,i}$ of subhalo $i$ follows the classical density-matching criterion $
\langle \rho_{{\rm gal},i}(\epsilon_i\, r^{\rm CDM}_{t,i})\rangle \simeq
\langle \rho_{\rm cluster}(r_{{\rm per},i})\rangle$, where $\epsilon_i$ captures order-unity scatter from formation histories, orbital diversity, and internal anisotropies \citep{Ghigna1998,TaylorBabul2001}. Calibrating $\epsilon_i \equiv r^{\rm dPIE}_{t,i}/r^{\rm CDM}_{t,i}$ using mass- and redshift-matched TNG-C analogs yields a broad distribution (central 97\% in $\epsilon_i\simeq 0.25$--1.18) with a median $\tilde{\epsilon}\simeq 0.48$ \citep{Chiang2026a}.

\begin{figure}
    \centering
    \includegraphics[width=0.45\textwidth]{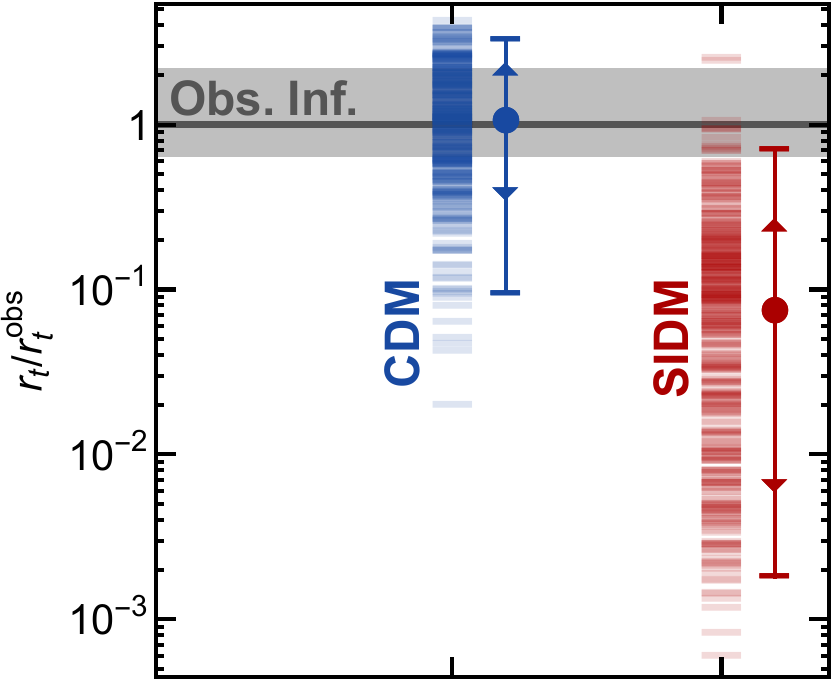}
\caption{Tidal radii from lensing-based observational inferences (gray line; the shading indicates conservative $5\sigma$ uncertainties), overlaid with individual CDM (blue horizontal dashes) and strongly collisional SIDM (red horizontal dashes; conservative upper bounds) predictions for all spectroscopically confirmed member galaxies in our sample. We also plot the respective medians (circles), central 68\% ranges (triangles), and central 95\% ranges (horizontal bars). Lensing inferences are fully statistically consistent with CDM and discrepant with SIDM.}
\label{fig:rtidal.pdf}
\end{figure}

For velocity-independent SIDM in the strongly collisional regime, ram-pressure-like mass stripping dominates, and the new truncation boundary is given by $\rho_{{\rm gal},i}(r^{\rm SIDM}_{t,i})\,\sigma_{{\rm gal},i}^2 \simeq
\rho_{\rm cluster}(r_{{\rm per},i})\,v_{{\rm per},i}^2$ \citep{FurlanettoLoeb2002}. This generally predicts much more compact radii than collisionless CDM. Using observed projected distances and line-of-sight velocities, \citet{Chiang2026a} adopt a conservative upper bound $r^{\rm SIDM}_{t,i} = \min\!\left(r^{\rm CDM}_{t,i},\, r^{\rm ram}_{t,i}\right)$ and compare these predictions to the lensing-inferred $r_t^{\rm obs}$ for all member galaxy subhalos across the eight clusters. Here, we show in \fref{fig:rtidal.pdf} the comparison for 362 member galaxy subhalos in our sample of three cluster lenses (\sref{sec:obs_cluster}).

The observed truncation radius distributions are fully consistent with CDM. In contrast, strongly collisional SIDM predictions are discrepant by orders of magnitude and remain well outside the observational $5\sigma$ band; projection effects, orbital uncertainties, and the calibrated scatter in $\epsilon_i$ cannot bridge the gap. Therefore, \citet{Chiang2026a} conclude that for the subhalo mass range contributing most strongly to GGSL, their outer structure is consistent with collisionless tidal stripping, excluding strongly collisional SIDM models capable of driving core collapse throughout the halo. Rather than quote a single exclusion value for $\sigma_{\rm SIDM}/m_{\rm SIDM}$, they emphasize that the mapping between the truncation radius and
cross section is highly nonlinear and requires a large number of simulated independent realizations to marginalize over unknown subhalo orbital distributions and infall histories.

In velocity-dependent SIDM models, self-interaction-driven evaporation becomes comparatively negligible \citep[e.g.,][]{Dooley2016,Zeng+2022}, but the cored central density profile can promote additional mass loss for inner subhalos in cluster environments \citep[e.g.,][]{Errani2023}, relative to the cuspy CDM benchmark. We similarly quantify this effect in \aref{app:sub_rtid}, by considering the velocity-dependent SIDM model of \citet{Nadler2023b}. Using the L-Cluster suite from SIDM Concerto \citep{Nadler2025arXiv250310748N}, we compare tidal radii for physically matched CDM--SIDM subhalo pairs. In this mildly collisional regime, subhalos are either CDM-like (20\%) or in a long-lived core phase (80\%), with collisionality far below that needed to resolve GGSL \citep{YangYu2021, Dutra2025}. Yet core formation can itself enhance tidal mass loss, up to complete disruption \citep[e.g.,][]{Errani2023}; for inner substructures, we show in \fref{fig:App_C} that this drives order-of-magnitude shifts in the ensemble-averaged tidal radius relative to CDM subhalos with NFW-like cusps. Overall, cluster subhalos behave as if DM self-interactions are negligible.

\begin{figure*}[ht!]
\centering
\includegraphics[width=0.85\textwidth]{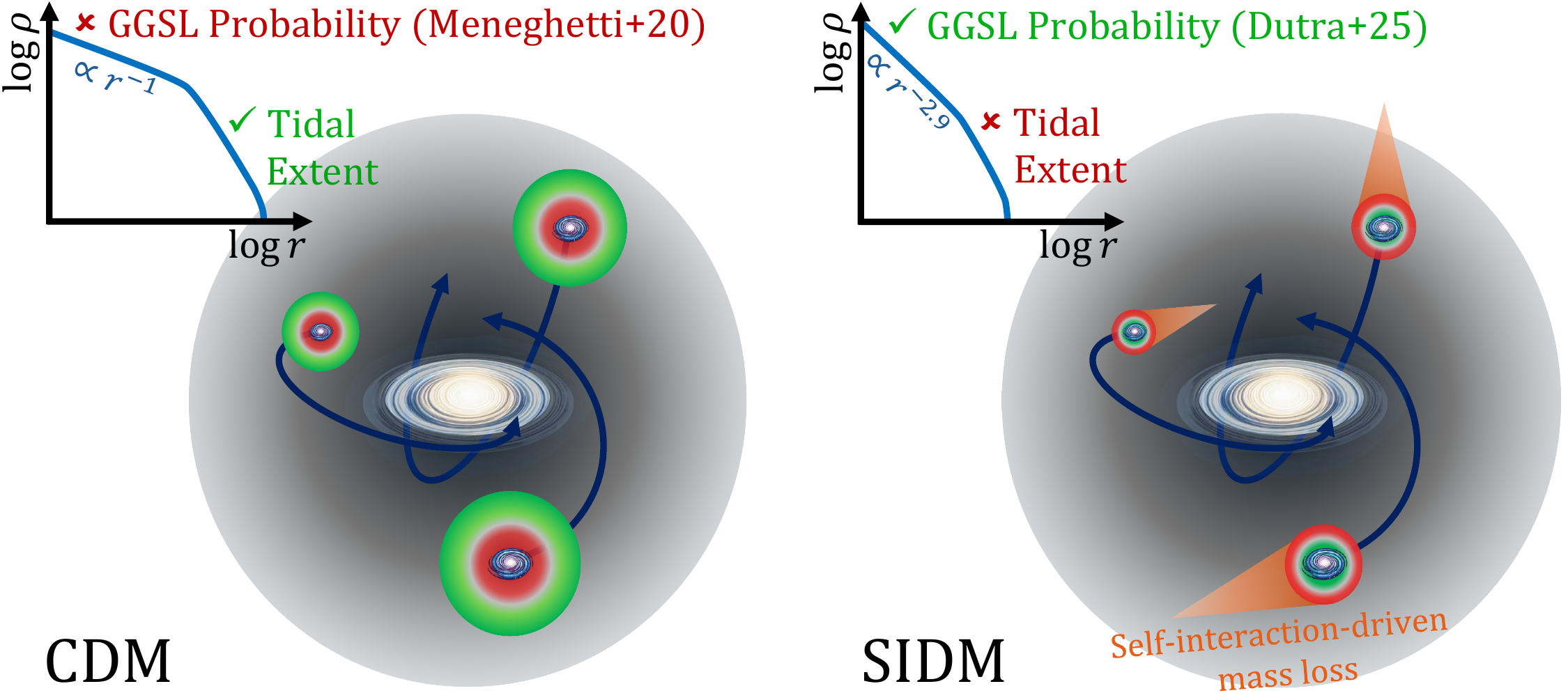}
\caption{Schematic illustrating the implications for CDM and SIDM models. Left: collisionless CDM matches subhalo tidal extents but severely underpredicts GGSL. Right: core-collapsed SIDM yields steep inner profiles that match GGSL, but this strongly collisional regime also results in more pronounced mass stripping that significantly reduces tidal extents, even after accounting for the slight expansion of the outskirts due to outward energy transfer during the core collapsing. The inset axes show the run of $\log\rho$ versus $\log r$ (not to scale).}
\label{fig:cluster_schematic}
\end{figure*}

Section~\ref{sec:ggsl_tension} and \ref{sec:trun_radii} therefore isolate two robust lensing constraints (Fig.~\ref{fig:cluster_schematic}). At large radii ($r\gtrsim50$ kpc), subhalo tidal extents and outskirts match collisionless CDM \citep{Chiang2026a}. At small radii ($r\lesssim10$ kpc), the GGSL probability requires much steeper inner slopes, $\gamma\gtrsim2.5$, than CDM+baryons produce \citep{Meneghetti2020,Meneghetti2022,Dutra2025,Tokayer2024}. Constant-cross-sectional SIDM can raise the GGSL rate via core collapse, but if interactions persist across the halo, they overstrip subhalos and shrink tidal radii. Velocity-dependent SIDM yields smoother radial trends tied to local velocity dispersion but can still reduce truncation radii during the core phase. If CDM is to be modified, these results point to models where self-interactions are \emph{inactive} in the outskirts yet \emph{active} only in dense collapsed central regions, so as not to incur additional self-interaction-driven mass loss---either via new dark-sector physics or a hybrid CDM+SIDM model. This is the first time that this inconsistency in expected self-interaction has been demonstrated in the dense cluster environment.

\section{Sources of uncertainty in our analysis}\label{sec:rt_Uncertainties}

In this section, we present a comprehensive assessment of the potential sources of uncertainties that could impact our results.
\begin{enumerate}[wide=0pt, nosep]
    \item \textit{Observational systematics:} The subsample analyzed in this work (\tref{tab:Cluster_Properties}) is part of a larger sample of eight independently and well-studied cluster lenses \citet{Chiang2026a}, which all have multiple independent ground- and space-based observations taken over a period spanning 25 yr, with data collected, reduced, and modeled by multiple independent groups \citep[e.g.,][]{LeBorgne1992AAS9587L, Smith2005MNRAS359417S, Newman2013ApJ76524N, Lotz2017ApJ83797L, Caminha2017AA600A90C} rendering identical systematics extremely unlikely. The $r_\text{t}$ analysis by \citet{Chiang2026a} finds consistent results across all clusters; our results---the SHMF agreement and inner abundance discrepancies---are also statistically consistent with independent analyses of A2744 \citep{Natarajan2017}.
    \item \textit{Lensing modeling:} The \Lenstool~parametric method robustly recovers subhalo properties in an unbiased fashion \citep{Meneghetti2017MNRAS4723177M}, with all independent parametric methods yielding excellent agreement for the total mass partitioned into subhalos \citep{Lotz2017ApJ83797L, Caminha2017AA607A93C, Caminha2019A&A632A36C, Natarajan2024}. This is one of the key outputs of the lensing analysis that we have utilized in this work. The distribution of mass within small-scale apertures is robustly obtained across different lensing modeling methodologies. Additionally, in this work, we select only subhalos hosting spectroscopically confirmed member galaxies, to guard against spurious detections and contaminating interlopers \citep{Ephremidze2025MNRAS5422610E}. The empirical scaling relations (\eref{eqn:E_Scaling_Relations}) are well motivated by established observations  \citep{Faber1976ApJ204668F, Natarajan:2002cw}, Bayesian-optimized \citep{Kneib1996ApJ471643K, Natarajan1997MNRAS287833N}, and verified for robustness \citep{Richard2010MNRAS404325R, Eichner2013ApJ774124E, Desprez2018MNRAS4792630D}.
    \item \textit{Numerical resolution and baryonic feedback in simulations:} Resolution-limited numerical artifacts remain a nontrivial concern for cosmological simulations, especially for substructures at small halocentric radii \citep{vandenBosch2018MNRAS4754066V, vandenBosch2018MNRAS4743043V, Martin2024, Chiang2026Universal}. Baryonic feedback also impacts the inner substructure survivability, with effects varying depending on specific subgrid implementations \citep[e.g.,][]{Despali2017, GarrisonKimmel2017}. To explore the robustness (see the details presented in \aref{app:num_rad_sub}), we injected subhalos into TNG-C analogs to match a DM-only, force-disruption-free benchmark. Performing 1 million bootstrap iterations showed that CDM predictions remain $5\sigma$--$40\sigma$ discrepant with the observed subhalos within $0.2 R_{200}$ (\fref{fig:App_B}). The GGSL discrepancy has been independently verified against resolution and feedback variations \citep{Meneghetti2020, Meneghetti2022, Meneghetti2023}. For tidal truncation radii, similar CDM--SIDM discrepancies are found by comparing matched subhalo pairs in DM-only zoom-in simulations (\fref{fig:App_C}), remaining consistent even with resolution-based quality cuts. These discrepancies we demonstrate are robust against numerical uncertainties.
\end{enumerate}

\section{Conclusions and Discussion}
\label{sec:conclusions}

Stress testing key substructure properties in observed lensing clusters with their simulated CDM analogs from the TNG-C suite, we find intriguing results that not all current observational constraints can be accounted for by collisionless DM. Our findings from the four independent physical diagnostics are summarized below.
\begin{enumerate}[wide=0pt, nosep]
\item{\textit{SHMFs:}} Reasonably good statistical agreement is found between the SHMFs derived from cluster lenses and those from simulated CDM TNG-C clusters. This agreement has been previously reported from other simulation suites like IllustrisTNG and The Three Hundred Project\citep{Natarajan2017,Meneghetti2023}.
\item{\textit{Outer subhalo structure:}} Lensing-inferred subhalo tidal truncation radii are in excellent agreement with collisionless CDM, ruling out any additional sizable contribution of self-interaction-induced mass loss on these scales of $\sim 10$\textendash$100\,{\rm kpc}$.
\item{\textit{Inner subhalo structure:}} The high efficiency of the observed GGSL requires significantly concentrated subhalo inner mass distributions with steeper-than-NFW central profiles, which are naturally produced only in the core-collapse regime of SIDM.
\item{\textit{Radial distribution of subhalos:}} The projected spatial distribution of the inner substructures in the observed massive clusters is highly discrepant with CDM. Simulated subhalos do not appear to adequately populate in the inner regions $R \leq 0.2\,R_{200}$ of clusters, where there is a significant concentration of observed bright cluster galaxies.
\end{enumerate}

Taken together, our results reveal that no single DM model currently appears capable of self-consistently reproducing the more detailed aspects of substructure demographics in observed massive clusters. Statistically, CDM matches the SHMF and the outer subhalo extents but cannot account for the inner lensing anomalies. The radial distribution of subhalos is also highly discrepant, with a marked absence of simulated subhalos at projected clustercentric radii $R \leq 0.2\,R_{200}$. On the flip side, while introducing strong self-interactions (SIDM and its variants) can account for the observed inner lensing efficiency, this would however inevitably also result in a more suppressed SHMF and an even stronger depletion of inner substructures relative to CDM, due to enhanced tidal mass loss by self-interaction-induced ram pressure stripping and core formation \citep[e.g.,][]{Nadler:2020ulu, Nadler2025arXiv250310748N}, rendering these other concordances in tension. Therefore, reconciling cluster-scale lensing data across all radii may require hybrid models in which DM behaves collisionlessly in the outskirts but exhibits effective self-interaction in the dense cores of subhalos. This newly inferred dichotomy, we conclude, highlights a critical challenge for the microphysics of DM.

For the galaxy-scale lens JVAS B1938+666 at $z_\text{spec} = 0.881$, \citet{Powell2025} and \citet{Vegetti2026} claim the detection of a dark, compact $\simeq 10^6~\Msun$ object whose mass profile can be described with a point-mass-like component---potentially arising from core-collapse SIDM, like the inner parts of the subhalos we reported in \citet{Dutra2025}---and an additional extended mass distribution with a radius of 139~pc. Independently, \citet{Yu2025arXiv251011006Y} also report consistent interpretation of this compact object. In contrast, we here make the case for core-collapse SIDM in the dense cluster environment.

In this Letter, informed by multiscale lensing analysis that effectively probes physical scales of 5\textendash100~kpc and substructure properties across multiple independent dense cluster environments, we present a new and fundamental paradox: the incompatibility of the nature of DM microphysics with the requirement to self-consistently account for the inner and outer properties of cluster subhalos. We demonstrate that while CDM severely underpredicts the lensing efficiency of the inner parts of subhalos constrained by GGSL observations, it can successfully explain the lensing-inferred outer tidal truncation radii of subhalos. This inherent lack of self-consistency is a real new flavor of crisis for CDM, distinct from the previously reported ones (\fref{fig:cluster_schematic}).

In summary, gravitational lensing offers a powerful, indispensable, and unique probe of DM theories. The further stress testing of DM models demands new, even-higher-resolution cosmological simulations and improved lensing data for larger cluster samples, which are expected imminently from EUCLID and the LSST Rubin Observatory. The anomalies we report here warrant further investigation, while robustly pointing the way to potentially new DM theories.

\begin{acknowledgments}
We acknowledge useful conversations with Frank van den Bosch and constructive comments from the anonymous referee. P.N. gratefully acknowledges funding from the Department of Energy grant DE-SC001766; support from the John Templeton Foundation via grant 126613; and support from NASA GLIMPSE JWST GO-03293.026 for this work. I.D. acknowledges support from NASA under award No. 80NSSC25K0311 under the NASA FINESST program. 

\end{acknowledgments}

\bibliography{refs}
\bibliographystyle{aasjournal}

\newpage
\appendix
\setcounter{figure}{0} 
\renewcommand{\thefigure}{A\arabic{figure}} 

\section{Lens Modeling}\label{app:lens_modeling}

The larger-scale halos as well as the subhalo population are modeled as self-similar dPIE mass distributions, whose 3D density profile $\rho_\text{dPIE}$, enclosed mass profile $M_\text{dPIE}$, and 2D surface density profile $\Sigma_\text{dPIE}$ are given by:
\begin{align}\label{eqn:rho_dPIE}
		&\rho_\text{dPIE}(r) \eee \frac{\rho_0}{\Big(1+\frac{r^2}{r^2_\text{core}}\Big)\Big(1+\frac{r^2}{r^2_\text{t}}\Big)},\\
        &M_\text{dPIE}(r) = 4\pi\rho_0\bigg(\frac{r_\text{core}^2 r_\text{t}^2\big[r_\text{t} \tan^{-1}\big(\frac{r}{r_\text{t}}\big)-r_\text{core} \tan^{-1}\big(\frac{r}{r_\text{core}}\big)\big]}{r_\text{t}^2-r_\text{core}^2}\bigg),\nonumber \\
		&\Sigma_\text{dPIE}(R) = \frac{\Sigma_0 r_\text{core}}{1-\big(\frac{r_\text{core}}{r_\text{t}}\big)}\bigg(\frac{1}{\sqrt{\smash[b]{r_\text{core}^2}+R^2}}-\frac{1}{\sqrt{\smash[b]{r_\text{t}^2}+R^2}}\bigg),\nonumber
\end{align}
where $r_\text{core}$ denotes the core radius, and $r_\text{t}$ denotes the tidal truncation radius of the subhalo. In this parameterization, the total mass $M_\text{dPIE}(r\rightarrow\infty) = 2\pi \Sigma_0 r_\text{core} r_\text{t}$ is finite.

For each subhalo, the normalization coefficients $\rho_0$ and $\Sigma_0$ are set uniquely by the effective velocity dispersion:
\begin{align}
\label{eqn:sigma_dPIE}
    \sigma_\text{dPIE} \eee \frac{4 G\pi \rho_0}{3}\frac{r_\text{core}^2 r_\text{t}^3}{(r_\text{t}-r_\text{core})(r_\text{t}+r_\text{core})^2} 
    = \frac{4G \Sigma_0}{3}\frac{r_\text{core} r_\text{t}^2}{r_\text{t}^2-r_\text{core}^2},
\end{align}
where $G$ denotes the gravitational constant. The numerical values of $\sigma_\text{dPIE}$ are derived for all substructures simultaneously, by optimizing the entire observed cluster lens images. It turns out that $\sigma_\text{dPIE}$ is related to the physically measured central velocity dispersion of each member galaxy by $\sigma_\text{dPIE}^2 = \frac{2}{3}\sigma_\text{gal}^2$, which is available for the member galaxies in the clusters studied here.

With the further assumption that light traces mass on cluster galaxy scales, the free parameters associated with individual subhalos are constrained by the empirical scaling relation of cluster member galaxies \citep{Natarajan:2002cw, Eliasdottir2007arXiv07105636E,Limousin2024}
\begin{align}
\label{eqn:E_Scaling_Relations}
	\begin{aligned}
        &\sigma_\text{dPIE} = \sigma_{\text{dPIE}*}\bigg(\frac{L}{\Ls}\bigg)^\alpha,\\
		&r_\text{t} = \rts\bigg(\frac{L}{\Ls}\bigg)^\beta,\\
	&r_\text{core} = r_{\text{core}*}\bigg(\frac{L}{\Ls}\bigg)^{1/2},
    \end{aligned}
\end{align}
where quantities with a star subscript denote the characteristic member galaxy properties, obtained by fitting a Schechter function to the luminosities of the hosted member galaxies \citep{Schechter1976ApJ203297S}. The values $\alpha = 0.25$ and $\beta=0.5$ corresponding to the Faber--Jackson relation \citep{Faber1976ApJ204668F} are favored for two of the best-fit cluster mass models of our sample, bar MACS~J1206, where we find best-fit values of $\alpha = 0.28$ and $\beta = 0.64$, respectively.

As noted above, the total mass of individual substructure scales as $M_\text{dPIE}(r\rightarrow\infty) = (9\sigma_{\text{dPIE}*}^2 r_{\text{t}*}/2G)(L/L_*)^{2\alpha+\beta}$, yielding a mass-to-light ratio that scales as $\Upsilon \propto(L/L_*)^{2\alpha+\beta-1}$. Physically, $2\alpha+\beta = 1$ corresponds to a system with a mass-independent mass-to-light ratio (although possible spatial dependence in $\Upsilon$ is still allowed during the Bayesian optimization); $2\alpha+\beta > 1$ indicates that brighter member galaxies exhibit larger $\Upsilon$.

\section{Exploring Numerical Uncertainties in Subhalo Abundance in Inner Regions}\label{app:num_rad_sub}

\begin{figure*}
    \centering
    \includegraphics[width=0.28\linewidth]{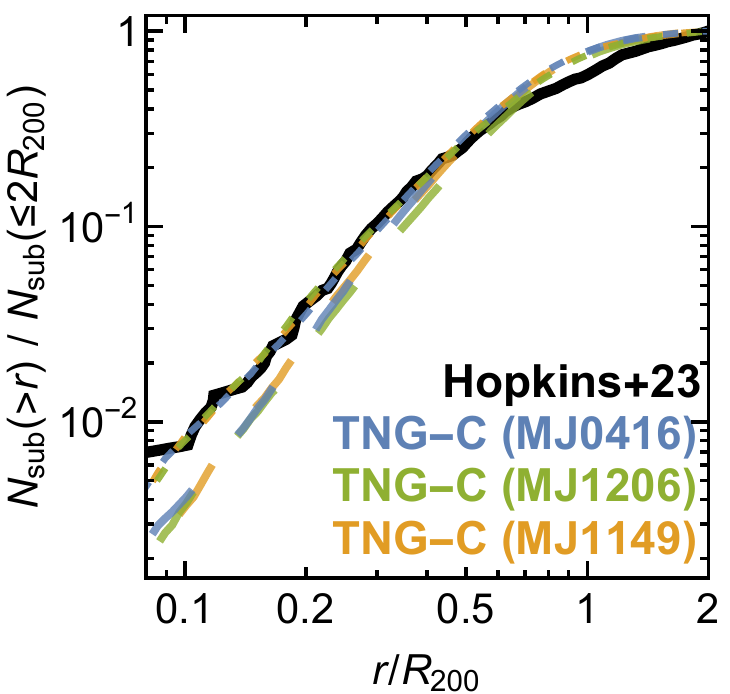}
    \includegraphics[width=0.695\linewidth]{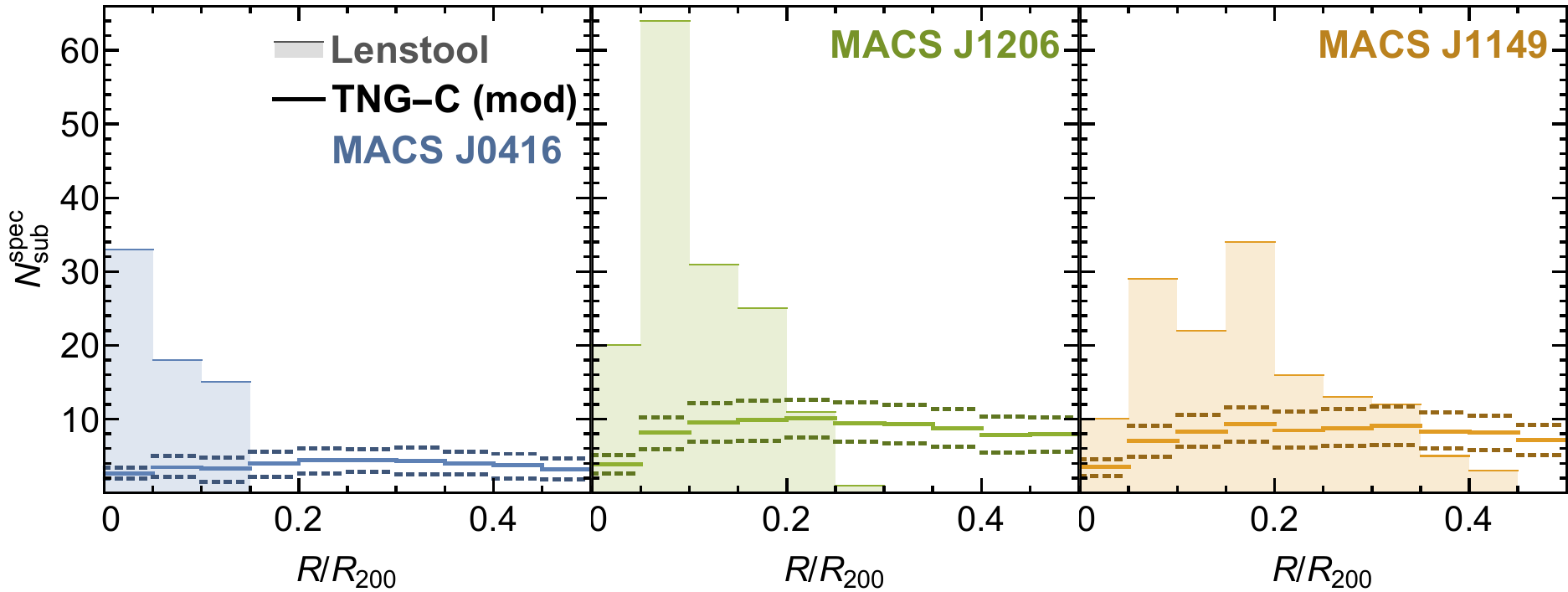}
    \caption{Left: 3D radial distribution of the subhalos in our TNG-C cluster analogs (dashed, color-coded as indicated) compared to the force-disruption-free benchmark taken from \citealt{Hopkins2023} (black solid). We show the cumulative number of all subhalos with more than 20 DM particles, normalized to the total number within the respective $2 R_{200}$. To explicitly account for the artificial disruption of the inner subhalos, we artificially ``inject'' subhalos within $0.5R_{200}$ by sampling weighted by the deficit within $0.5R_{200}$ between the TNG-C and benchmark cumulative distributions (see the text for details). Three random examples of such modified subhalo radial distributions (dotted) are plotted that well match the benchmark distribution. Right: projected radial distribution of the subhalos hosting member galaxies in observed clusters (color-shaded, as in the top panel of \fref{fig:combined_figures}) and in simulated analogs with injected inner subhalos. The solid (dotted) lines show the mean ($1\sigma$ confidence interval) inferred from 1 million bootstrapping iterations. Within the projected radius of $\lesssim 0.2R_{200}$, individual radial bins remain $5\sigma$\textendash$40\sigma$ discrepant with the observed abundance, even after explicitly accounting for numerical uncertainties due to artificial disruption.}
	\label{fig:App_B}
\end{figure*}

As demonstrated in \citet{Chiang2026Universal}, with the kiloparsec-scale force resolution adopted in TNG-C, subhalos become nontrivially impacted by force-resolution-limited artifacts when halocentric pericenter distances drop below $\lesssim 0.5 R_{200}$ and are almost entirely force-unresolved if the pericenter distances lie below $\lesssim 0.2 R_{200}$. To assess how this numerical uncertainty impacts the projected luminosity-selected subhalo radial distribution and hence the robustness of our results in \sref{sec:radial_dist}, we perform the following illustrative exercise, by ``manually injecting'' additional subhalos into the existing TNG-C simulated subhalo catalog. We outline below how we explicitly account for the numerically disrupted inner subhalo population.

Here, we adopt as a benchmark the cumulative distribution function (CDF) of the 3D subhalo radial distribution of an LMC-mass host from a DM-only cosmological zoom-in simulation presented in \citet{Hopkins2023}. Specifically, we adopt the run with tidal-radius-based adaptive gravitational softening (the $\xi = 2$ case in the top panel of Figure~9 therein), which ensures the inner subhalos remain properly force-resolved and free from force-resolution-limited numerical runaway disruption \citep{Chiang2026Universal}\footnote{This benchmark still suffers from mass-resolution-limited artifacts, especially for severely particle-limited subhalos near the 20 particle threshold. However, as noted in \citet{Chiang2026Universal}, mass-resolution-limited artifacts cause an \emph{unbiased} spread in the bound mass fraction (i.e., some subhalos lose more and some less than the convergence values) and thus should not statistically reduce the subhalo number counts, provided the subhalos are properly force-resolved.}. This benchmark represents a conservative upper limit on the normalized abundance of inner subhalos, as the addition of baryons and baryonic feedback can exacerbate the depletion of inner substructures in cluster environments (\citealp{Chua2017, Despali2017}; see however \citealp{Haggar2021}). Furthermore, in DM-only simulations, the normalized abundance of inner substructure in an LMC-sized host has been found to be, on average, comparable to that in massive-cluster-scale hosts \citep{Gao2004} or even slightly higher \citep{Nadler2023}\footnote{It is possible that part of the difference in the inner subhalo abundance observed in the left panel of \fref{fig:App_B} is driven by differences in host halo concentrations (and not just by artificial disruption alone), with typical values of $c \sim 15$ for LMC-mass hosts and $c \sim 4$ for massive clusters. Granted this, the LMC-scale benchmark again serves as a highly conservative (or optimistic) choice for inner substructure abundance.}. Last, the subhalo properties of \citet{Hopkins2023} were computed using \textsc{Rockstar} \citep{Behroozi2013}, which accesses the full 6D phase-space information and shows demonstrable improvement in subhalo identification compared to methods that only use 3D spatial information, such as \textsc{SubFind} \citep[e.g.,][]{vandenBosch2016MNRAS}. Taken together, this benchmark CDF of the subhalo radial distribution provides a ``highly optimistic'' calibration of the inner subhalo abundance that we incorporate into TNG-C analogs.

As the subhalo radial distribution of \citet{Hopkins2023} was computed for all subhalos above 20 particles and lying within $2R_{200}$, we follow the same practice and compute the corresponding CDFs of the TNG-C analogs (averaged over the five best-mass-matched analogs) per observed cluster, shown as the dashed curves in the left panel of \fref{fig:App_B}. Next, we minimize the difference between these TNG-C CDFs and that of \citet{Hopkins2023}, by ``injecting'' subhalos within a 3D halocentric radius of $R\leq 0.5R_{200}$ (accounting for those artificially disrupted due to inadequate force resolution), sampling from a distribution weighted by the difference in CDFs. We stop the injection when the resulting CDF has the same area as that of \citet{Hopkins2023} when integrating from the smallest radial bin to $0.5R_{200}$; three such examples are shown as the dotted curves in the left panel of \fref{fig:App_B}. Empirically, we find that this matching requires us to increase the subhalo abundance within $0.5R_{200}$ by 12\%\textendash30\%, consistent with the expected deficit due to artificial disruption in cosmological simulations reported by \citet{Green2021MNRAS5034075G}.

Next, the physical properties (the subhalo mass and magnitude of the associated member galaxy) of these injected objects are randomly resampled from the existing subhalo population of each TNG-C analog, given the consistency between the simulated and observed SHMFs (\fref{fig:SHMF.pdf})\footnote{While our injection procedure samples $V$-band magnitudes independent of radius, empirical studies show that satellites in massive clusters exhibit luminosity segregation \citep[e.g.,][]{Adami1998, Lin2004}, where more luminous member galaxies concentrate more strongly toward the cluster core, due to dynamical friction and merging \citep{FuscoFemiano1998}. Accounting for this empirical distribution would likely increase the predicted inner density, thereby lessening the reported statistical discrepancy between CDM and observations. However, because resolution-limited artifacts already preferentially disrupt low-mass and thus low-luminosity systems in simulations \citep[see also][]{Brainerd2021}, it is unclear whether sampling from a modified distribution is more physically justified than resampling directly from the existing simulated distribution.}. Therefore, some of these objects can be completely dark (i.e., have no star particles). For each injected subhalo, we combine its 3D halocentric radius with a unit vector uniformly sampled from a unit sphere to define its 3D spatial position. We then take the entire subhalo population (originally simulated + injected) and project it along three random orthogonal directions. Finally, we select subhalos based on the $V$-band magnitude of their associated member galaxies, following the same observational selection function described in \sref{sec:sim_tngc}.

We perform such statistical bootstrap resampling iterations to obtain $10^6$ independent projections on the simulated analogs for each observed cluster. The right panel of \fref{fig:App_B} shows the average distribution and the respective $1\sigma$ confidence interval. Even after explicitly accounting for the missing satellite population due to inadequate force-resolution-driven artificial disruption, the substructure abundance within a projected radius of $0.2 R_{200}$ as predicted by CDM is still statistically discrepant by $5\sigma$\textendash$40\sigma$ from the observations of massive-cluster lenses.

\section{Subhalo Truncation Radii in Velocity-dependent SIDM
Models}\label{app:sub_rtid}

In \sref{sec:trun_radii}, we demonstrated that observational inferences based on strong gravitational lensing are statistically consistent with CDM-like subhalos that experience only collisionless tidal mass loss while maintaining the NFW-like central density cusp. In contrast, the core-collapsed SIDM subhalos---as required to reconcile GGSL (\sref{sec:ggsl_tension})---yield tidal truncation radii that are discrepant by an order of magnitude below the observational inferences due to pronounced ram pressure stripping in the strongly collisional regime (\fref{fig:rtidal.pdf}). While the analysis of \citet{Chiang2026a} was done under the implicit assumption of a constant (i.e., velocity-independent) cross section, such that most subhalos within the GGSL-pertinent mass scale $M_\text{sub} \sim 5\times 10^{9}$\textendash$10^{12}~\Msun$ in massive clusters are core-collapsed \citep{Dutra2025}, SIDM models with velocity-dependent cross sections can also drive observable deviations in tidal mass loss from the CDM predictions and be tested with the tidal-truncation-radius-based diagnostic. In particular, self-interaction-driven constant-density cores can be long-lasting in the mildly collisional regime, which renders subhalos much more susceptible to collisionless tidal mass loss and physical core disruption \citep[e.g.,][]{Errani2023}.
 
Here, we demonstrate that a similar discrepancy in the ensemble-averaged tidal truncation radius, as reported in \fref{fig:rtidal.pdf} relative to the CDM predictions, is also found in SIDM subhalos, even under the assumption of a velocity-dependent cross section in the mildly collisional regime (i.e., most subhalos are in the core phase and none are in the core-collapsed phase), where now the discrepancy is predominantly driven by stronger tidal mass stripping, due to the presence of self-interaction-driven central thermalized density cores. We directly compare the physically matched subhalos extracted from the SIDM Concerto suite presented in \citet{Nadler2025arXiv250310748N}, which comprises CDM and velocity-dependent SIDM DM-only cosmological zoom-in simulations from identical initial conditions. In particular, we select their L-Cluster run ($M_{200} = 1.6\times~10^{14}~\Msun$) at a particle resolution of $m_\text{DM} = 2.7\times10^7~\Msun$ and a fixed gravitational softening length of 0.857 kpc; at $z = 0$, the subhalos span the mass range of $M_\text{sub}\simeq 10^8$\textendash$10^{13}~\Msun$ and extend out to a 3D halocentric radius of $1.89 R_{200}$. The SIDM run assumes a velocity-dependent cross section with the Rutherford-like parameterization $\rmd \sigma/\rmd \cos\theta \eee \sigma_0 w^4/{2[w^2+v^2\sin^2(\theta/2)]^2}$ \citep{Ibe2010, Yang2022}, with $\sigma_0/m_{\rm SIDM} = 147.1$~cm$^2$g$^{-1}$ and $w = 120$~km~s$^{-1}$. Under this choice, the effective cross section at massive-cluster scale is $\sigma/m_{\rm SIDM}(M_{200} \sim 10^{15}~\Msun)\sim10^{-2}$~cm$^2$g$^{-1}$ and monotonically increases until $M_{200} \sim 10^{10}~\Msun$, then plateaus at a cross section of $147.1$~cm$^2$g$^{-1}$. In the L-Cluster run, about 20\% of the SIDM subhalos remain CDM-like, 80\% are in the ``core'' phase, and none are in the core-collapsed phase.

To match CDM subhalos and their respective SIDM counterparts in formation and infall history, we first perform cross-simulation matching using the \textsc{Rockstar} + \textsc{consistent-trees} (RCT; \citealp{Behroozi2013, Behroozi2013b}) catalogs, where subhalos are tracked across time and assigned merger-tree branch identifiers based on their pre-infall trajectories. We then project these \textsc{Rockstar}-level matches into the \textsc{Symfind} \citep{Mansfield2024} catalogs, by using the merger-tree branch index as a persistent identity key. In the symlib/Concerto pipeline \citep{Kong2025}\footnote{\url{https://github.com/DemaoK/Concerto/tree/main/symlib}}, \textsc{Symfind} inherits these branch identifiers from the RCT merger tree, allowing each matched branch to be mapped to its corresponding \textsc{Symfind} entry. We have verified the robustness of this cross-simulation identification by the consistency in the matched subhalo peak masses, mass assembly histories, and orbital trajectories prior to and shortly after accretion. In constructing the final matched subhalo sample, we exclude all pairs where the CDM subhalo has ill-measured entry values for the instantaneous tidal radius $r_\text{tid}$, where \textsc{Symfind} adopts the definition of \citet{King.62} that accounts for both the extended mass distributions of the host and subhalo as well as the centrifugal force from the subhalo’s orbital motion. In cases where the SIDM subhalo has undergone physical core disruption prior to $z = 0$ but the CDM counterpart still has a well-measured $r_\text{t}$, we adopt $r_\text{tid} = 0$ for that SIDM subhalo, as recorded in the \textsc{Symfind} catalog.

\begin{figure*}
    \centering
    \includegraphics[width=0.98\linewidth]{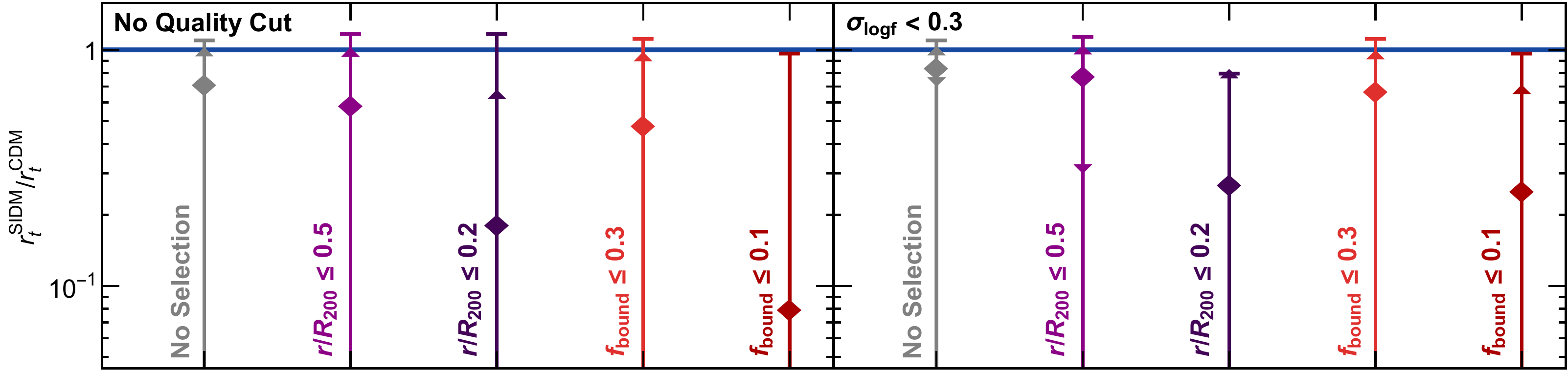}
    \caption{Ratios of subhalo tidal radii computed from CDM--SIDM-matched pairs of the L-Cluster run of the SIDM Concerto suite \citep{Nadler2025arXiv250310748N} that assumes a Rutherford-like velocity-dependent cross-section model (see the text for details). We show the mean (diamonds), central 68\% (triangles), and central 95\% (horizontal bars) ranges of each distribution under no subhalo selection (gray) and as selected by instantaneous halocentric radius (purple) or bound mass fraction (red). This comparison is done for the entire CDM–SIDM-matched subhalo sample (left) and for the subsample filtered by requiring the mass-resolution-limited numerical scatter in the bound mass fraction to be less than 0.3 dex (right; \eref{eqn:sigma_fbound}). Even in the mildly collisional regime or with velocity-dependent SIDM models, self-interaction-driven core formation leads to substantially higher tidal mass loss and tidal truncation radii statistically discrepant from CDM predictions, by up to an order of magnitude for inner and/or heavily stripped subhalos.}
	\label{fig:App_C}
\end{figure*}

The final sample comprises 645 such CDM--SIDM pairs. Importantly, the observed subhalo sample revealed by the strong lensing that enables the tidal truncation radius analysis (\fref{fig:rtidal.pdf}) lies within projected clustercentric distances of $0.2$–$0.5R_{200}$ (see \fref{fig:combined_figures}) and thus on average experiences more pronounced tidal mass loss than the entire subhalo population \citep[e.g.,][]{vandenBosch2016}. We perform additional selection on this sample, based on either the CDM subhalo instantaneous 3D orbital distance $r/R_{200}$ or bound mass fraction $f_\text{bound} \eee M_\text{sub}/M_\text{peak}$, where $M_\text{sub}$ is defined as the total bound mass of the subhalo at $z=0$ provided in the \textsc{Symfind} catalog. The left panel of \fref{fig:App_C} compares the distribution of the ratios of the CDM--SIDM-matched subhalo tidal radii under no selection (gray) with these aforementioned selections (color-coded, as indicated). Without any selection, the mean difference $r^\text{SIDM}_\text{t}/r^\text{CDM}_\text{t} = 0.73$ is within the observational inference uncertainties, but the discrepancy clearly increases with decreasing $r/R_{200}$ and $f_\text{bound}$. In particular, for subhalos lying within $0.2 r/R_{200}$ or with $f_\text{bound}\leq 0.1$, the ensemble-averaged SIDM subhalo tidal truncation radius falls below the CDM predictions by an order of magnitude.

Last, to also assess the possible impact of mass-resolution-limited numerical artifacts on this analysis, we apply a quality cut based on the realization-to-realization scatter in $f_\text{bound}$. In particular, based on the universal numerical scatter track of \citet{Chiang2026Universal} \begin{align} \label{eqn:sigma_fbound} \sigma_{\log f} = 1.8 N_\text{par}^{-0.5} f^{-0.6}_\text{bound} (1-f^{0.5}_\text{bound})^{0.8}, \end{align} where $N_\text{par} = M_\text{sub}/m_\text{DM}$, we keep all matched pairs whose CDM subhalo has $\sigma_{\log f} < 0.3$ (i.e., discreteness-noise-driven uncertainties in $f_\text{bound}$ less than 0.3~dex). This subsample comprises 508 matched pairs, with mostly inner low-mass and heavily stripped substructures removed in this process. We repeat the same $r/R_{200}$- and $f_\text{bound}$-based selections and compare the resulting $r^\text{SIDM}_\text{t}/r^\text{CDM}_\text{t}$ distributions in the right panel of \fref{fig:App_C}. The overall trend remains qualitatively unchanged; the more heavily stripped or inner subhalo subsample still shows near order-of-magnitude discrepancy from the CDM predictions. This analysis demonstrates that the tidal truncation radii of inner subhalos in massive lensing clusters constitute a powerful discriminator of DM self-interaction, even in the mildly collisional regime or for velocity-dependent SIDM models.

Before closing, we note that the present analysis relies on a direct object-by-object comparison of CDM and SIDM subhalo properties in a DM-only context. While this is a limitation, the inclusion of baryons introduces competing effects, whose net impact on the CDM--SIDM discrepancy is uncertain. On one hand, baryons can steepen the central density slopes via adiabatic contraction in CDM \citep[e.g.,][]{Gnedin2004} and accelerate the core collapse in SIDM \citep{Zhong2023}. Conversely, stellar and AGN feedback are expected to effectively lower the central density slopes \citep[e.g.,][]{Pontzen2012MNRAS4213464P, DiCintio2014, Tollet2016} within the mass range of the cluster subhalos considered here (\fref{fig:SHMF.pdf}). The net outcome is further complicated by the still-debated impact of baryons on CDM subhalo survival in the cluster environment \citep{Chua2017, Despali2017, Haggar2021}. However, we note that the order-of-magnitude discrepancy found for inner and heavily stripped subhalos suggests that our qualitative conclusions are unlikely to be fully negated by baryonic effects. A full assessment of baryonic effects on this CDM--SIDM discrepancy would require dedicated hydrodynamic simulations, which is beyond the scope of this work and deferred to future studies.

\end{document}